\documentclass[aps,prfluids,groupedaddress,amsmath,amssymb,longbibliography]{revtex4-1}

\usepackage{amsmath}%
\usepackage{MnSymbol}%
\usepackage{wasysym}%
\usepackage{graphicx}
\usepackage{dcolumn}
\usepackage{bm}
\usepackage[mathlines]{lineno}
\usepackage{color}

\begin{document}

\preprint{APS/123-QED}

%
%


\title{On the controversial turbulent Schmidt number value in particle-laden boundary layer flows}


%
%





\author{J. Chauchat}
\affiliation{%
Univ. Grenoble Alpes, CNRS, Grenoble INP*, LEGI, 38000 Grenoble, France* Institute of Engineering Univ. Grenoble Alpes}%
\author{D. Hurther}%
\affiliation{%
Univ. Grenoble Alpes, CNRS, Grenoble INP*, LEGI, 38000 Grenoble, France* Institute of Engineering Univ. Grenoble Alpes}%
\author{T. Revil-Baudard}
\affiliation{%
Univ. Grenoble Alpes, CNRS, Grenoble INP*, LEGI, 38000 Grenoble, France* Institute of Engineering Univ. Grenoble Alpes}%
\author{Z. Cheng}
\affiliation{%
Civil and Environmental Engineering, Center for Applied Coastal Research, University of Delaware, Newark, DE 19711, USA\\
Now at Convergent Science Inc., Madison, WI 53719, USA 
}%
\author{T.-J. Hsu}
\affiliation{%
Civil and Environmental Engineering, Center for Applied Coastal Research, University of Delaware, Newark, DE 19711, USA
}%
 \email{julien.chauchat@univ-grenoble-alpes.fr}

\begin{abstract}

One of the most enigmatic science question concerning inertial particle transport by a turbulent boundary layer flow is the value of the turbulent Schmidt number defined as the ratio of turbulent eddy viscosity to particle concentration diffusivity. Using direct acoustic measurement of turbulent particle flux profile, and two-phase flow turbulence-resolving numerical simulation, it is demonstrated that turbulent dispersion of particles is reduced rather than enhanced as predicted by many existing literature models. The explanation lies in the misleading assumption of settling velocity in quiescent water to estimate the turbulent particle diffusivity, while direct measurements and simulations of turbulent particle flux support the occurrence of settling retardation. The analysis presented herein suggests that the value of the turbulent Schmidt number is always larger than unity with values between 3 and 4 based on the directly measured turbulent particle flux. The observed settling reduction can not be explained by the well-known hindrance effects related to particle concentration. This effect seems to be related to turbulence-particle interactions and correlates more with the Stokes number. At last, the new parameters, namely the turbulent Schmidt number higher than unity, modified von K\'arm\'an constant, and settling retardation, are successfully tested for the modeling of particle concentration profile using the well-known Rouse formulation. This result suggests that new parametrizations are possible to reduce the degree of empiricism to predict suspended particle transport by a boundary layer flow.

\end{abstract}

\maketitle

%
%
\section{Introduction}

The transport of solid particles by a turbulent fluid flow is a key  process in geophysical and industrial two-phase flows such as material and food processing, pneumatic transport, fluidized beds, slurry flows or sediment transport. In all these situations, interactions between particles and fluid turbulence are key physical processes that contribute to the dynamics of the system. Among the consequences induced by the presence of particles, the modification of fluid turbulence and the turbulent dispersion of particles are crucial mechanisms that are not fully understood \cite{Balachandar2010}. The complexity arises when particles are inertial, \textit{i.e.} when the particle response time $\tau_p$ is larger than a representative turbulent eddy time-scale $\tau_f$, leading to Stokes number ${\rm St}=\tau_p/\tau_f$ greater than unity. Inertial particles act as a local filter for the turbulent kinetic energy spectrum due to their not fully-correlated movements with turbulent eddies of similar or smaller size than the particle itself. This problem has been studied in many different flow configurations including Homogeneous Isotropic Turbulence, jets and turbulent wall-bounded flows.  However, further complexity arises under gravitational acceleration.

In sediment transport, under intense flows corresponding to high fluid bed shear stress relative to the buoyant weight of the particles,  the so-called Shields number $\theta$,  strong erosions from the underlying sediment bed are observed and a significant amount of particles are transported in suspension by fluid turbulent eddies \cite{garcia2008}. The coupling between the sediment flux and the underlying bed topography controls the large-scale morphological evolution of rivers, estuaries and coastal nearshore zones \cite{aagard2010}. In most of these situations, the transported particles are made of inorganic non-cohesive sand grains with Stokes number values well above unity, \textit{i.e.} inertial particles. Sediment transport is a truly multi-scale process coupling fine scale turbulence-particle interactions (O($\mu m$) - O(ms)) with the morphological evolution of natural systems at large scales (O($km$) - O(yrs)). Therefore, elucidation of the governing interaction processes associated with the transport of inertial particles by a turbulent boundary layer flow is a major scientific issue.   

The quantity of particular interest is the stream-wise particle flux defined as the integral over the flow depth of the product between the local particle velocity and particle concentration. Consequently, it is necessary to understand the key processes that controls the velocity and concentration profiles. On the one hand, the presence of particles may significantly modify turbulence \cite[e.g.][]{vanoni1975,kiger2002,revil-baudard2015,cheng2018}. This is often characterized by a modification of the von K\'arm\'an constant $\kappa$ in the classical logarithmic law-of-the-wall. On the other hand, the turbulent diffusivity of particle concentration $\epsilon_\phi$ may significantly differ from the fluid eddy-viscosity $\nu^t$ \cite{lyn2008}. These two quantities are the key ingredients to predict the particle flux. Under fully-developped turbulent flow conditions and by assuming a local balance between the gravity-driven settling flux $W_s \langle\phi\rangle$ and the upward Reynolds particle flux   $\langle w'\phi' \rangle$, the particle phase mass conservation equation reduces to:
\begin{equation}
\label{MassBalance}
-W_s \langle \phi \rangle + \langle w'\phi' \rangle=0.
\end{equation}
In this equation, $W_s$ is the particle settling velocity and $\langle w'\phi' \rangle$ is the wall-normal Reynolds particle flux. In the later, $w'$ stands for the velocity fluctuations and $\phi'$ for the concentration fluctuations defined with respect to their  respective mean values where $\langle \rangle$ represents a temporal averaging operator. \citet{rouse1938} proposed to model the Reynolds particle flux using a Fickian gradient diffusion model:
\begin{equation}
\label{vertturbflux}
\langle w'\phi' \rangle = - \epsilon_\phi \dfrac{d \langle \phi \rangle}{d z},
\end{equation}
where $\epsilon_\phi$, the particle concentration turbulent diffusivity, is linked to the eddy-viscosity through the turbulent Schmidt number defined as ${\rm S_c}=\nu^t/\epsilon_\phi$. Using a Prandtl's mixing length model with a free surface correction for the eddy-viscosity $l_m=\kappa z (1-z/H_f)^{1/2}$, where  $\kappa$ is the von K\'arm\'an constant, $z$ is the wall-normal distance and $H_f$ is the water depth above the particle bed and by further assuming that particle settling velocity $W_s$ can be approximated by its value in quiescent water $W^0_s$,  \citet{rouse1938} derived an analytical solution for the mean particle concentration profile as:
\begin{equation}
\label{RouseProfile}
 \dfrac{\phi}{\phi_r} = \left( \dfrac{H_f-z}{z} \dfrac{z_r}{H_f-z_r}\right)^{Ro},
\end{equation}
where $\phi_r=\phi(z_r)$ is a reference concentration taken at an elevation $z_r$ above the particle bed and located within the suspension layer, \textit{i.e.} where $\phi<0.08$, and ${\rm Ro=S_c} W^0_s/\kappa u_*$ is the Rouse number corresponding to the slope of the particle concentration profile. In this definition, $u_*$ represents the bed friction velocity. Using measured concentration profile and the estimated bed friction velocity and particle settling velocity, the Schmidt number can be readily estimated using the Rouse profile. 

Discrepancies between a measured concentration profile and the Rouse profile have been attributed to a modification of the particle diffusivity compared with the eddy viscosity, leading to values of the turbulent Schmidt number ${\rm S_c}$ different from unity. Because this classical method estimates the Schmidt number by  best fitting the measured concentration profile with the Rouse profile and leaving ${\rm S_c}$ as the only free parameter, the resulting ${\rm S_c}$ is a depth-averaged value noted as ${\rm \overline{S_c}}$ in the following. Figure 1a summarizes several experimental values reported in the literature for ${\rm \overline{S_c}}$ plotted as a function of $W_s/u_*$ \cite{lyn2008}. The first observation that can be made is the large scatter of the data with values lower and higher than unity. Values lower (higher) than unity can be interpreted as an enhanced (reduced) dispersion of particles compared to fluid momentum. \citet{vanrijn1984b} suggested that turbulent dispersion of inertial particles is enhanced due to centrifugal force which tends to throw particles out of small-scale turbulent vortexes and proposed an empirical model written as:
\begin{equation}
\label{SchmidtVanRijn}
{\rm \overline{S_c}} = \left[ 1+ 2 \left(\dfrac{W_s^0}{u_*}\right)^2 \right]^{-1}.
\end{equation} 
While this model (solid line in Figure 1a) is widely used, it does not match the laboratory data of \citet{lyn1988}.

The turbulent Schmidt number can also be obtained locally provided that the concentration profile is measured with enough accuracy to resolve the sharp wall-normal gradients near the flow bed. Figure 1b shows a summary of existing data for locally estimated Schmidt number as a function of the normalized depth $z/H_f$. Again ${\rm S_c}$ shows large scatter with values ranging from as low as 0.2 to as large as 2.5. In the experiments reported by \citet{lyn1988} and \citet{barton1955}, only the mean concentration profile was measured and so ${\rm S_c}$ is obtained by using both the mass balance and the gradient diffusion model (Eqs. \ref{MassBalance}-\ref{vertturbflux}) and therefore under the assumption that $W_s=W_s^0$. In \citet{cellino1999}, using the novel acoustic particle flux profiler of \citet{shen1999}, the authors measured directly the turbulent particle flux and hence only Eq. \ref{vertturbflux} is used to calculate ${\rm S_c}$ directly. The mean concentration profile used to estimate ${\rm S_c}$ was obtained from an iso-kinetic flow suction system with a much lower vertical resolution than the direct measurement of the Reynolds particle flux \cite{CellinoPhD}. This introduces a source of uncertainty in their results especially in the near-bed region where the concentration gradient is very high. Despite this source of measurement uncertainty, it is interesting to point out that almost all their data (circles in 1b) suggest ${\rm S_c}>1$. 

The potential reasons for the discrepancies in the Schmidt number values reported in the literature are (i) the assumption of using particle settling velocity in quiescent water and (ii) the assumption of the von K\'arm\'an constant value being identical to the clear water value ($\kappa=0.41$). Concerning the first assumption, several publications have reported that the settling velocity of individual particles could be modified in turbulent flows \cite{murray1970,good2014,cuthbertson2007,kawanisi2008,zhou2009,akutina2020,mora2021}. However, no  consensus has been reached on the effect of turbulence on the settling velocity of individual particles as some studies have found settling velocity enhancement while others observed settling retardation \cite[e.g.][]{nielsen1992,kawanisi2008,dey2019}. Regarding the second assumption, in particle-laden flows,  von K\'arm\'an constant reduction has been reported  \cite[e.g.][]{vanoni1941,revil-baudard2015,revil-baudard2016}. Two physical mechanisms have been proposed to explain this reduction, density stratification \cite{villaret95} and/or turbulent drag work \cite{hsu2003,cheng2018}. 
The fact that the ratio ${\rm S_c}/\kappa$ appears in the Rouse number definition could explain some of the observed discrepancies.  In order to address this question, it is mandatory to measure concurrently and with sufficient accuracy the velocity and the concentration profiles to allow for a direct estimate of the von K\'arm\'an constant and of the turbulent Schmidt number.  This ability represents a major measurement challenge under such energetic particle-laden flow conditions.

 To the best of our knowledge, the first measurements reported in the literature showing turbulent fluxes, mean concentration and velocity profiles are those of \citet{shen1999} using the Acoustic Particle Flux Profiler. Over the past ten years, we have intensively developed and improved this technology as the Acoustic Concentration and Velocity Profiler (ACVP) offering a unique wide-band multi-frequency capability for simultaneous vertical profiles of velocity and particle concentration at turbulence resolving scales ($\Delta z \approx 3$ mm ; $f = 78$ Hz) \cite{hurther2011a,hurther2011b,thorne2014,naqshband2017,fromant2018a,fromant2019}. We successfully applied this technology to study intense sediment transport processes in the so-called sheet-flow regime in \citet{revil-baudard2015}. This unique data-set allows for the first time to investigate  the relationship between  Reynolds particle flux, Reynolds stresses, and velocity and particle concentration gradients in order to shed new light on the turbulent Schmidt number value.

Over the last two decades, significant progress has been made on the numerical modeling of sediment transport by using two-phase flow approaches  \cite[e.g.][]{chauchat2008,jha2010,revil-baudard2013,lee2016,cheng2016,chauchat2018}. Recently, we have performed the first turbulence-resolving two-fluid simulation \cite{cheng2018} that has been validated against experimental data from \citet{revil-baudard2015}. In this approach, the most energetic turbulent flow scales are directly resolved as well as the particle dynamics allowing the majority of turbulent particle fluxes to be resolved. 


In this contribution, using the recent high-resolution experimental data-set of \citet{revil-baudard2015} and turbulence-resolving two-phase flow simulations of \citet{cheng2018}, we are able to provide a physical explanation for the contradictory findings in the literature. 
\begin{figure}[ht]
\begin{center}
\includegraphics[width=0.6\textwidth]{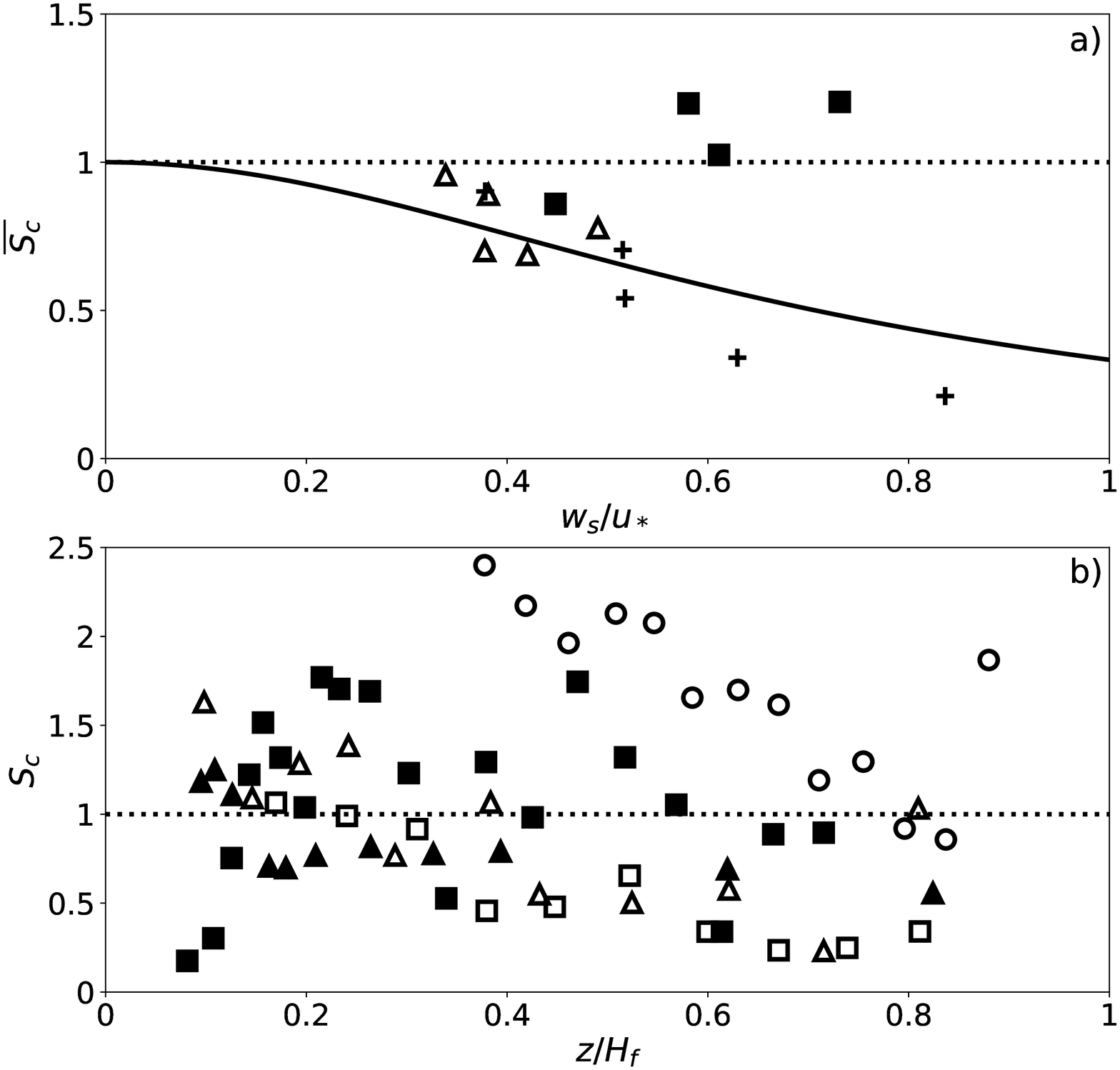}
\caption{Experimental data for the turbulent Schmidt number ${\rm S_c}$ from \citet{lyn2008}: a)  Averaged values of ${\rm S_c}$ over $0.1< z_/H_f<0.5$ as a function of $W_s/u_*$: $\triangle$ \citet{barton1955}, $+$ \citet{coleman1970}, $\blacksquare$ \citet{lyn1988}, the black solid line corresponds to \cite{vanrijn1984b} model (Eq. \ref{SchmidtVanRijn}) ; b) Local estimates of ${\rm S_c}$ for uniform flow over a plane equilibrium bed as a function of relative distance, $z/H_f$, from the bed: $\circ$ \citet{cellino1999}, $\blacksquare$ \citet{lyn1988}, $\blacktriangle$ \citet{lyn1988}, $\square$ \citet{barton1955}, $\triangle$ \citet{barton1955}. }
\label{fig0}
\end{center}
\end{figure} 

\section{Material and method}

\subsection{Sediment transport experiments}
\label{appendixMeasurements}

In \citet{revil-baudard2015} sediment transport experiments have been carried out in the LEGI/ENSE3 tilting flume. The mobile bed experiment is denoted as $RB15$ and corresponds to a Shields number of $\theta=u_*^2 / [(\rho_p/\rho_f-1) g d_p]=0.4$ where $\rho^f$ and $\rho^s$ are fluid and particle densities, $g$ is gravity acceleration and $d_p$ is the particle diameter ; and a suspension number of $W_s^0/u_*=1.2$  (see table \ref{table_run_properties}). For this experiment the slope has been set to $S_0=0.5\%$. The experimental condition has been realized $N=11$ times for $RB15$ in order to verify the reproducibility of the experiment and to perform ensemble averaging.  The experimental  parameters are presented in table \ref{table_run_properties}.

The ACVP technology was used to measure the vertical profiles, from the non-moving particle bed up to the free-surface, of the two velocity components $u(z),w(z)$ and of the particle volumetric concentration $\phi(z)$ with vertical spatial resolution of $\Delta z=3$mm at 78 Hz (see \citet{thorne2014} for details on acoustic measurement method and  \citet{revil-baudard2015,revil-baudard2016} for details on the specific ACVP settings applied in the datasets used herein).

\begin{table}[ht]
\caption{Summary of experimental parameters. Particle diameter ($d_p$), density ratio ($\rho_p/\rho_f$), settling velocity ($W_s^0$), friction velocity ($u_*$),  suspension number ($W_s^0/u_*$), bulk Reynolds number (${\rm Re_b} =U H_f/\nu^f$), with $\nu^f=10^{-6} m^2/s$, particulate Reynolds number (${\rm Re_p}=W_s^0 d_p /\nu^f$), Stokes number based on the Kolmogorov time scale (${\rm \overline{St_\eta}}=\tau_p / \overline{\tau_\eta}$ with $\tau_p$  the particle response time and $\overline{\tau_\eta}=\overline{\eta}^2/\nu^f$ the Kolmogorov time-scale with $\overline{\eta}=((\nu^f/u_*)^3 d_p )^{1/4}$) and Stokes number based on the viscous time scale (${\rm St^+}=\tau_p / \tau_v$ with $\tau_v=\nu^f/u_*^2$) . }
\begin{center}
\begin{tabular}{cccccccccccccccccc}
\hline
\hline
Name            &           Reference                 & $d_p$     & $\rho^s/\rho^f$ & $W_s^0$   &$u_*$      & $W_s/u_*$ & ${\rm Re_b}$               & ${\rm Re_p}$ & ${\rm \overline{St_\eta}}$ & ${\rm St^+}$ \\
                     &                                              &($mm$)    &                                        & ($cm/s$)   & ($cm/s$) &                           &      ($\times 10^4$)    &    &   &\\
\hline
$RB15$ & \citet{revil-baudard2015}         & $3$         &  $1.19$            & $6.2$         & $5.0$          & $1.2$               &$9$  & 186 & 17.4 & 178.8\\
$SL99\_1$ &\citet{shen1999} & $0.13$   & $2.65$             & $1.2$         & $4.8$         & $0.25$               &$26 $  & 1.3 & 1.9    &  2.3\\
$SL99\_2$ &\citet{shen1999} & $0.13$   & $2.65$             & $1.2$         & $5.2$           & $0.23$               &$29 $  &1.3 & 2.2 & 2.7 \\
 \hline 
 \hline
\end{tabular}
\end{center}
\label{table_run_properties}
\end{table}

A specific experimental protocol has been applied to obtain steady and uniform flow conditions in the absence of a particle recirculating facility. First, the time period $t\in[t_1-t_2]$ for which the flow is uniform is identified based on the vertical linearity of the mean turbulent shear-stress profile. Then the local and instantaneous velocity, concentration and particle flux measurements are temporally averaged over $\Delta t=t_2-t_1$ and over the $N$ realizations of the same experiment to guarantee statistically converged velocities, concentration and particle flux profiles. 
\begin{equation}
\displaystyle \langle A \rangle^S(z)= \frac{1}{N} \sum_{i=1}^N  \left(\frac{1}{\Delta t} \int_{t_1}^{t_2} A_i(t,z)  dt \right).
\label{U}
\end{equation}
The operator $\langle - \rangle^S$ refers to the superficial averaging, and $A$ is a given measured quantity. The averaging time window is  $\Delta t=6$s.

In order to discuss and confirm the validity of the results obtained for RB15 experiments, the experimental dataset of \citet{shen1999} will be used for comparison and referred to as runs SL99\_1 and SL99\_2 in the following. To the best of our knowledge, it is the only existing experimental dataset providing the required direct measurements of time-resolved vertical profiles of two components velocity, particle concentration and fluxes in similar type of particle-laden flows. The SL99 experiments were carried out in similar open-channel flow conditions regarding the hydrodynamic regimes (fully turbulent, fully hydraulically rough, uniform, steady and subcritical flows). Run SL99\_2 is somewhat more energetic than SL99\_1 in terms of hydraulic power but similar regarding the hydrodynamic and sediment transport regimes. The particles (sand grains of well-sorted unimodal size distribution) used in the two SL99 experiments explain the important differences seen in Table I for the suspension number, particle Reynolds and Stokes numbers values. These values reveal a  fully suspension dominated particle-laden flow (with negligible bedload transport). This is due to the less inertial sand grains having much smaller size but higher density compared with the PMMA particles used in RB15. Finally, another important point is that for all considered particle-laden flows, transport capacity was fully established. This implies that concentration effects are at their maximal possible level for all considered fluid-particle interaction processes.
%
\subsection{Two-fluid turbulence-resolving simulations}
\label{appendixModel}

In the two-fluid turbulence-resolving or Large Eddy Simulations (LES) model, the large-scale turbulent structures are directly computed from filtered Navier-Stokes equations while the small-scale motions, occurring at spatial scales smaller than the filter size, are accounted for using sub-grid closures. 
In this numerical approach, both the fluid phase and the particle phase are modeled as a continuum. A Favre filtering approach is used for the scale separation, and the filtered continuity equations read as, 
\begin{equation}
\frac{\partial ({1-\hat{\phi}})}{\partial {t}} + \frac{\partial{({1-\hat{\phi}})\hat{u}^f_i}}{\partial{x_i}}=0,
\label{eq_conti_fl}
\end{equation}
\begin{equation}
\frac{\partial{\hat{\phi}}}{\partial {t}} + \frac{\partial{\hat{\phi}\hat{u}^s_i}}{\partial{x_i}}=0,
\label{eq_conti_sed}
\end{equation}
\noindent
where $\hat{\phi}$ is the filtered particle volumetric concentration, $\hat{u}^f_i,\hat{u}^s_i$ are the filtered fluid and particle velocities, and $i=1, 2, 3$ represents streamwise ($x$), spanwise ($y$) and vertical ($z$) components, respectively.

The filtered momentum equations for fluid phase and particle phase are written as:
\begin{widetext}

\begin{eqnarray}
\frac{\partial {\rho^f}({1-\hat{\phi}})\hat{u}^f_i}{\partial {t}} + \frac{\partial{{\rho^f}({1-\hat{\phi}})\hat{u}^f_i\hat{u}^f_j}}{\partial{x_j}} & = & -(1-\hat{\phi})\frac{\partial{\hat{p}^f}}{\partial{x_i}}+\frac{\partial{(1-\hat{\phi})(\hat{\tau}^f_{ij}+\hat{\tau}^{f,sgs}_{ij})}}{\partial{x_j}} \nonumber \\
&& + \rho^f(1-\hat{\phi})g_i+\hat{M}^{fs}_i,
\label{eq_mom_fl}
\end{eqnarray}
\begin{equation}
\frac{\partial {\rho^s}\hat{\phi}\hat{u}^s_i}{\partial {t}} + \frac{\partial{{\rho_p}\hat{\phi}\hat{u}^s_i\hat{u}^s_j}}{\partial{x_j}}=-\hat{\phi}\frac{\partial{\hat{p}^f}}{\partial{x_i}}+\frac{\partial{\hat{\phi}\hat{\tau}^{s,sgs}_{ij}}}{\partial{x_j}}-\frac{\partial{\hat{p}^s}}{\partial{x_i}}+\frac{\partial{\hat{\tau}^s_{ij}}}{\partial{x_j}} + \rho^s\hat{\phi}g_i-\hat{M}^{fs}_i
\label{eq_mom_sed}
\end{equation}
\end{widetext}

where $g_i$ is the gravitational acceleration and $\hat{p}^f$ is the fluid pressure. $\hat{\tau}^f_{ij}$ and $\hat{\tau}^{f,sgs}_{ij}$ are the fluid (molecular) viscous stress and subgrid stress associated with the unresolved turbulent motions. In analogy to the fluid phase, the unresolved particle motions due to turbulence are taken into account by the subgrid stress, $\hat{\tau}^{s,sgs}_{ij}$. These subgrid stresses are modeled using a dynamic Smagorinsky subgrid closure \citep{Germano1991,Lilly1992}. The particle pressure $\hat{p}^s$ and particle stress $\hat{\tau}^s_{ij}$ due to intergranular interactions are modeled by the kinetic theory of granular flows \citep{Ding1990} and phenomenological closure of contact stresses \citep{Srivastava2003}. $\hat{M}^{fs}_i$ represents the filtered drag force between fluid phase and particle phase, which is only composed of  the resolved part in the present work: 
\begin{equation}
\hat{M}^{fs}_i=-\widehat{\phi\beta{u^r_i}}\approx-\beta\widehat{{\phi}u^r_i}=-\beta\hat{\phi}\hat{u}^r_i ,
\label{eq_mtransf}
\end{equation}
where $\hat{u}^r_i = \hat{u}^f_i-\hat{u}^s_i$ is the resolved relative velocity between fluid phase and particle phase. Contrary to \citet{cheng2018} the sub-grid drag model is ignored in the present work.

The details of the numerical configuration can be found in \citet{chauchatGMD2017,cheng2018} and the major parameters are summarized in table \ref{TableSimus}). Spatial averaging over the two statistically homogeneous $x$ and $y$ directions and time averaging are applied to the simulation results over a duration of about 60 eddy turn over time $T^e=H_f/U=0.175$ s to obtain ensemble-averaged flow statistics. 

\begin{table}[ht]
\caption{Numerical parameters used for the two-phase flow LES sheet flow simulations.}
\begin{center}
\begin{tabular}{cccccccccc}
\hline
\hline
        run name          &  N cells         &  $\Delta t$ & $\Delta_x$ &  $\Delta_y$  &     $\Delta_z^{min}$ & $\Delta_z^{max}$  \\
                                 &   (in millions) &  (s)            & (mm)      &     (mm)      &     (mm)      &     (mm)      \\
\hline
 $LES$  &  29.2  &$ 2 \times 10^{-4}$ &   1.65 & 1.65 &  0.4  & 2.2     \\
\hline
\hline
\end{tabular}
\end{center}
\label{TableSimus}
\end{table}%

\section{Results and discussion}

In the following, our unique high-resolution experimental dataset, denoted as RB15, and two-fluid turbulence-resolving simulation results, denoted as two-fluid LES, are analyzed to infer whether ${\rm S_c}>1$ or ${\rm S_c<1}$ (see table \ref{table_run_properties}). 

\begin{figure*}[t]
\centerline{
\includegraphics[width=\textwidth]{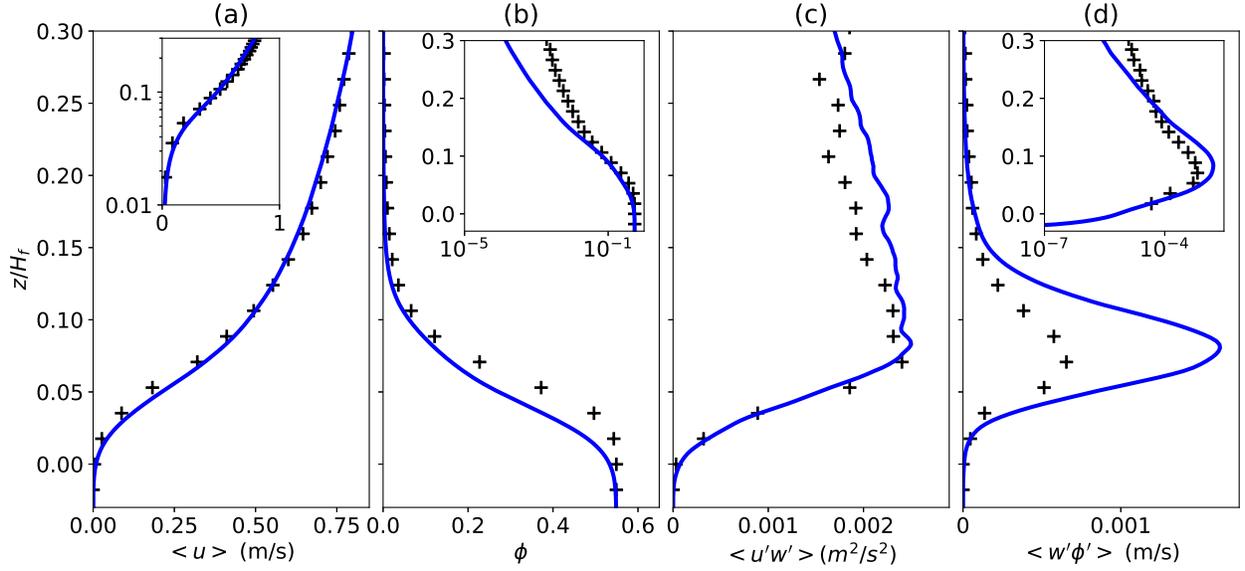}
}
\caption{Turbulence averaged vertical profiles of mean velocity $\langle u \rangle$ (a), mean concentration  $\langle \phi \rangle$ (b), the Reynolds shear stress $\langle u' w' \rangle$ (c) and the wall-normal particle  flux $\langle w' \phi' \rangle$ (d) for measured data (+) and  $LES$ results (-----).}
\label{fig1}
\end{figure*}

The averaged profiles of velocity, concentration, Reynolds shear stress and particle flux corresponding to this configuration are presented in figure \ref{fig1}. Measured Reynolds stress $\langle u' w' \rangle$ (symbols in Figure \ref{fig1}c) peaks near the top of the sheet flow layer (particle volumetric concentration about 0.08) and shows the expected upward-decaying linear profile valid for steady, uniform channel flows. Within the sheet flow layer, Reynolds stress decays sharply towards the bed. Measured particle flux in the wall-normal direction $\langle w' \phi' \rangle$ also peaks at the same vertical location (Figure \ref{fig1}d). The results obtained using two-fluid LES are compared with measured data from \citet{revil-baudard2015} in figure \ref{fig1}. In terms of velocity and concentration profiles (Figure \ref{fig1}a and b) the numerical results are in good agreement with the measured data although particle concentration in the dilute suspension region is underestimated. The numerical results reproduce measured Reynolds stress $\langle u' w' \rangle$ very well. However, the vertical turbulent particle flux $\langle w' \phi' \rangle$ is over-predicted (resp. under-predicted) in the near bed region (resp. in the outer layer). This probably explains the discrepancies on the concentration profile observed in the two-fluid LES results. These discrepancies are still an open question. It is important to recall here that the maximum of the wall-normal turbulent flux is located in the near-bed region where the particle concentration is high, \textit{i.e.} $\phi \in [0.1; 0.55] $. At such volumetric concentrations a strong interplay between turbulence-particle interactions and particle-particle interactions, the so-called four-way coupling, is taking place. It is highly possible that some of the closures of the two-fluid model requires modifications to better reproduce the measurements. This issue deserves further investigations but more experimental data are needed to guide the development of the two-fluid model.  

\begin{figure*}[t]
\centerline{
\includegraphics[width=\textwidth]{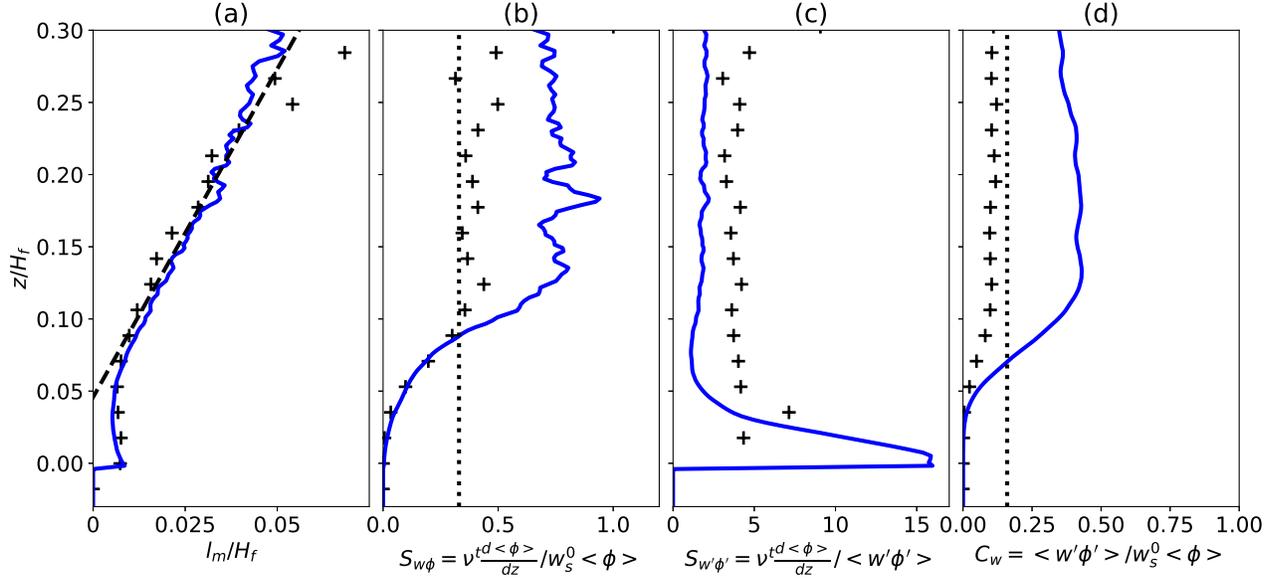}
}
\caption{Vertical profiles of (a)  mixing length, (b) Schmidt number defined as $S_{w\phi}=-\nu^t d \langle \phi \rangle / d z / W_s^0 \langle \phi \rangle$ Eq. (\ref{SchmidtNumber1}) (dotted line : empirical model \citet{vanrijn1984b} $S_{w\phi}=0.33$ Eq. \ref{SchmidtVanRijn}), (c) Schmidt number based on the resolved Reynolds flux $S_{w'\phi'}=\nu^t d \langle \phi \rangle / d z / \langle w'\phi' \rangle $ Eq. (\ref{SchmidtNumber2}) (d) dimensionless settling velocity: ${\rm C_w}= \langle w'\phi' \rangle / W_s^0 \langle \phi \rangle$ Eq. (\ref{EffectiveSettling}) (dotted line :  empirical formula \citet{akutina2020} ${\rm C_w}=0.16$ Eq. \ref{EffectiveSettling}). Measured data RB15 (black +) and two-fluid \ LES results (blue -----).}
\label{fig2}
\end{figure*} 

In order to avoid any uncertainties associated with a modification of the von K\'arm\'an constant, the eddy viscosity is calculated directly as:
 \begin{equation}
\label{EddyViscosity}
\nu^t = \frac{\vert \langle u'w'\rangle\vert}{\frac{d \langle u \rangle }{ d z}}.
\end{equation} 
To better illustrate the change in the von K\'arm\'an constant value, figure \ref{fig2}a presents the turbulent mixing length $l_m = \sqrt{\vert\langle u'w'\rangle\vert}/ d \langle u \rangle / d z$). The dashed black line in the figure represents the best-fit of the analytical model $l_m=\kappa (z-z_d)$ where the slope of the line is the von K\'arm\'an constant. As pointed out in \citet{revil-baudard2015}, the von K\'arm\'an constant value obtained in the experiment is $\kappa\approx0.22$, a significantly lower value than 0.41 expected for clear water flow conditions. The two-fluid LES reproduces this reduction very well. The modification of the von K\'arm\'an "constant" value has significant fundamental consequences for boundary layer modeling in particle-laden flows. Furthermore, as already mentioned, the von K\'arm\'an "constant" appears at the denominator of the Rouse number and severely impacts suspended particle transport modeling. 

\begin{figure*}[t]
\centerline{
\includegraphics[width=\textwidth]{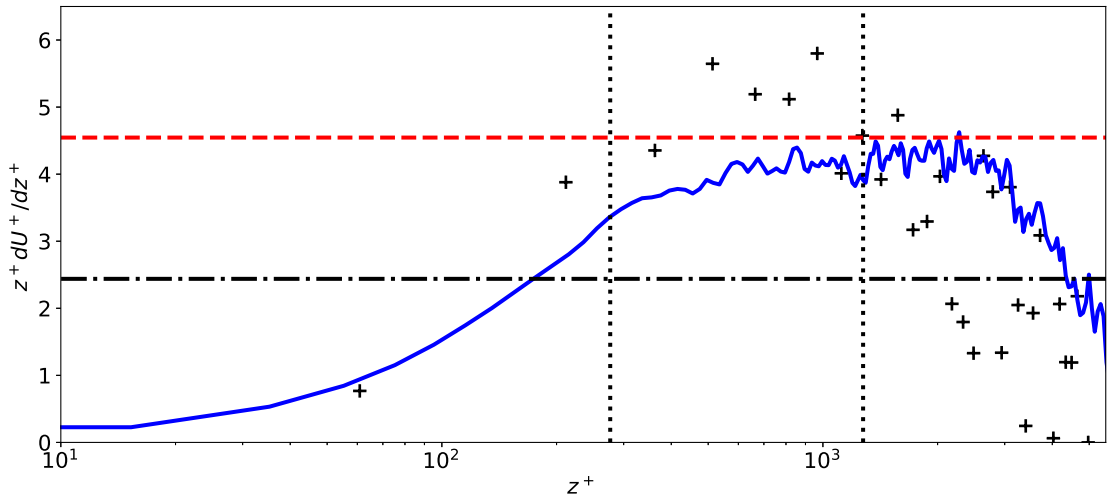}
}
\caption{Profile of $z^+ dU^+/dz^+$ as a function of $z^+$ in semi-logarithmic scale. Measured data RB15 (black +) and two-fluid \ LES results (blue -----) ; the horizontal black dash-dotted line represents $1/0.41$ and the horizontal red dashed line represents $1/0.22$ ; the vertical black dotted lines represents the range of the log-layer $3\sqrt{{\rm Re_\tau}}<z^+<0.15 {\rm Re_{\tau}}$ with ${\rm Re_\tau}=u_* H_f/\nu^f=8,500$.  }
\label{fig1bis}
\end{figure*} 
In order to confirm the existence of the logarithmic layer in this particle-laden flow configuration, the classical diagnostic for single phase flows is presented. It consists in plotting the product $z^+ dU^+/dz^+$ as a function of $z^+$ \cite[e.g.][]{smits2011}. A first difficulty to overcome in particle-laden flows over erodible beds is the definition of the origin for $z^+$. In this work, we suggest to use $z_d$, the virtual origin of the mixing length:
$$z^+=\dfrac{(z-z_d) u_*}{\nu^f}$$
Figure \ref{fig1bis} shows the $z^+ dU^+/dz^+$ as a function of $z^+$ in semi-logarithmic scale. Both RB15 data and the two-fluid LES exhibit a plateau confirming the existence of the log-layer. The range of $z^+$ over which the log layer exists does not correspond exactly with the classical values for single phase flows: $3\sqrt{{\rm Re_\tau}}<z^+<0.15 {\rm Re_{\tau}}$ with ${\rm Re_\tau}=u_* H_f/\nu^f=8,500$ (vertical dotted lines in figure \ref{fig1bis}). This difference may be attributed either to the definition of the origin of $z^+$ or to turbulence modifications induced by the presence of particles. Nevertheless, the existence of a plateau confirm that the flow velocity follow a logarithmic profile. The range retained to estimate the von K\'arm\'an constant is $z/H_f \in [0.05;0.275]$ corresponding to $z^+ \in [0; 2\times10^3]$ with our definition of $z^+$.

The classical method to estimate the turbulent Schmidt number is indirect and uses Eq. \ref{MassBalance} by assuming a local balance between the gradient-diffusion model (Eq. (\ref{vertturbflux})) and the gravity-driven settling flux $W_s^0\langle\phi\rangle$ with the particle settling velocity estimated by its value in quiescent water: 
\begin{equation}
\label{SchmidtNumber1}
S_{w\phi} = \frac{\nu^t }{W_s^0 \langle \phi \rangle} \Big \vert\frac{d \langle \phi \rangle }{d z}\Big \vert.
\end{equation}
This quantity will be denoted as $S_{w\phi}$ and is shown in figure \ref{fig2}b. For $z/H_f\gtrsim0.1$, $S_{w\phi}$ is more or less constant with a value below unity $S_{w\phi}\approx 0.4$ for RB15 data and $S_{w\phi}\approx0.7$ for two-fluid LES results. The model proposed by \citet{vanrijn1984b} (see Eq. \ref{SchmidtVanRijn}) gives a value of ${\rm \overline{S_c}}=0.3$  which is very close to the experimental value.

The direct method is based on the measured or simulated eddy viscosity and Reynolds particle flux (Eq. \ref{vertturbflux}). Using these quantities, the  turbulent Schmidt number can be calculated without using Eq. \ref{MassBalance} as:
\begin{equation}
\label{SchmidtNumber2}
S_{w'\phi'} = \frac{\nu^t }{\langle w'\phi' \rangle} \Big \vert\frac{d \langle \phi \rangle }{ d z}\Big \vert.
\end{equation}
This quantity will be denoted as $S_{w'\phi'}$ and is shown in figure \ref{fig2}c. The profiles are almost constant for $z/H_f\ge 0.1$, and the measured $S_{w'\phi'}$ is about 4 in RB15 while the two-fluid LES  confirms a value of $S_{w'\phi'}$ well above unity, even-though slightly smaller. As mentioned above, measurement of the Reynolds particle flux remains a particularly challenging task which, to the best of the authors knowledge, has only been provided to date in \citet{shen1999} and \citet{revil-baudard2015} for similar type of particle-laden open-channel flows. Our analysis demonstrated that the estimation of Schmidt number is highly dependent on the methodology used.


In a multi-phase flow approach, it is well-known that inertial particles cannot respond instantaneously to the wide range of fluid velocity fluctuations that exists in a highly turbulent flow. The particle dynamics may be significantly different from that of the fluid parcels \cite{Balachandar2010}. Particle inertia is estimated using the particle Stokes number ${\rm St_\eta}=\tau_p / \tau_\eta$ where $\tau_\eta$ is the Kolmogorov time-scale associated with the smallest turbulent eddies. Inertial particles correspond to ${\rm St_\eta}>1$ which is the case for the configurations investigated herein (see table \ref{table_run_properties}). This supports the observation that turbulent dispersion of particles should be less efficient than turbulent mixing of fluid momentum \textit{i.e.} $S_{w'\phi'}>1$.
Therefore, the indirect estimate of turbulent Schmidt number lower than unity $S_{w\phi}<1$ is hypothesized to be due to reduction of settling velocity in the settling flux. In order to prove this point, a dimensionless  settling velocity ${\rm C_w}$ is deduced from the mass balance equation, as: 
\begin{equation}
\label{EffectiveSettling}
{\rm C_w} = \frac{\langle w'\phi' \rangle}{ \langle \phi \rangle W_s^0}.
\end{equation}
This quantity is plotted in figure \ref{fig2}d which shows an almost constant value in the dilute suspended-load layer ($z/H_f\ge 0.1$) for both RB15  and two-fluid LES results. The measured quantity shows that the settling velocity is about ten times smaller than the one in quiescent water \textit{i.e.} ${\rm C_w}\approx0.1$. Recently \citet{akutina2020}, using the same PMMA particles as the one used in RB15 experiments, have measured a strong settling retardation for individual particles in a turbulence water column under quasi-homogeneous and isotropic turbulent flow conditions. In these experiments, the turbulence was generated by facing pairs of vertical grids oscillating horizontally. The measured settling velocity of individual particles has been observed to drastically decrease with increasing turbulence intensity $w_{rms}$. Using five different particle types covering a wide range of size, density and settling velocity in quiescent water, the authors have shown that all the data points for the dimensionless settling velocity ${\rm C_w}$ as a function of the dimensionless  turbulent intensity $w_{rms}/W_s^0$ collapse on a master curve given by: 
\begin{equation}
{\rm C_w} = 1 - 1.3 \dfrac{w_{rms}}{W_s^0},
\label{W_s_equ}
\end{equation}
for $0.2<w_{rms}/W_s^0<0.6$. In the present open-channel flow experiment, the vertical turbulent intensity can be estimated as $w_{rms} \approx 0.8 u_* \approx 4$ cm/s and Eq. (\ref{W_s_equ}) gives a value of ${\rm C_w}=0.16$. This value is very close to the one shown in figure \ref{fig2}d. 


 The two-fluid LES also predicts settling retardation (${\rm C_w} \approx0.4$) even though the reduction is less strong than in the experiments. Despite the discrepancy in magnitude, both numerical and measured values strongly support the hypothesis that settling retardation is responsible for the indirectly measured Schmidt number value smaller than unity. 
 
\begin{figure*}[t]
\centerline{
\includegraphics[width=\textwidth]{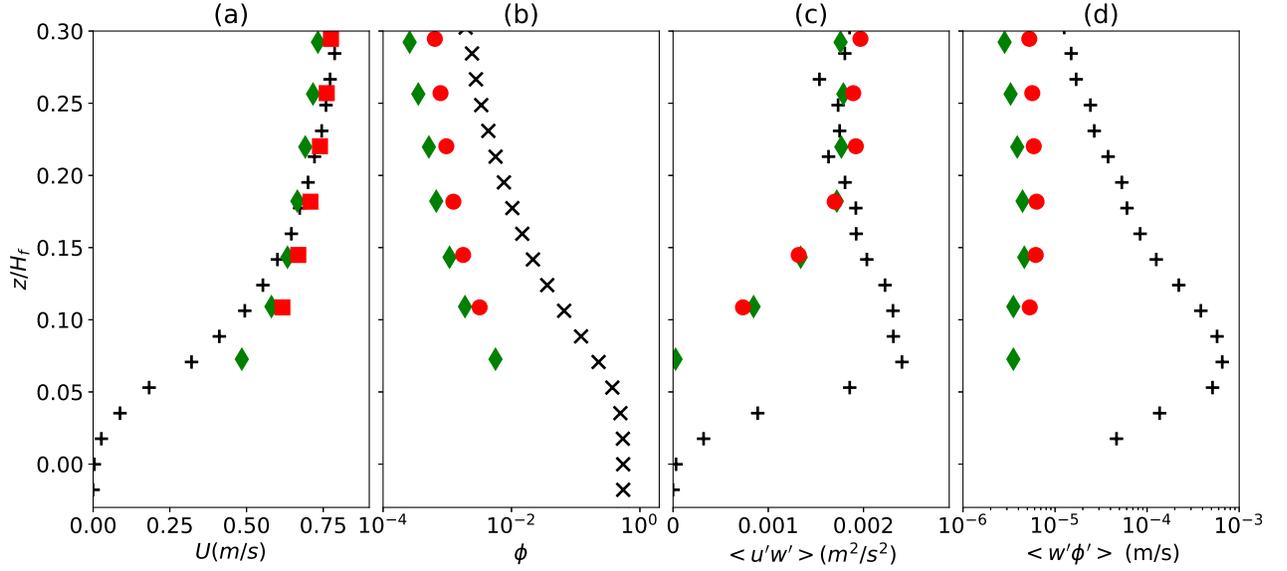}
}
\caption{Comparison of RB15 data with $SL99\_1$ (green diamond $\diamond$) and $SL99\_2$ (red circle $\circ$)  data from \citet{shen1999}  reported for two different hydrodynamic conditions. Same panels as in figure \ref{fig1}. }
\label{fig3}
\end{figure*}
\begin{figure*}[t]
\centerline{
\includegraphics[width=\textwidth]{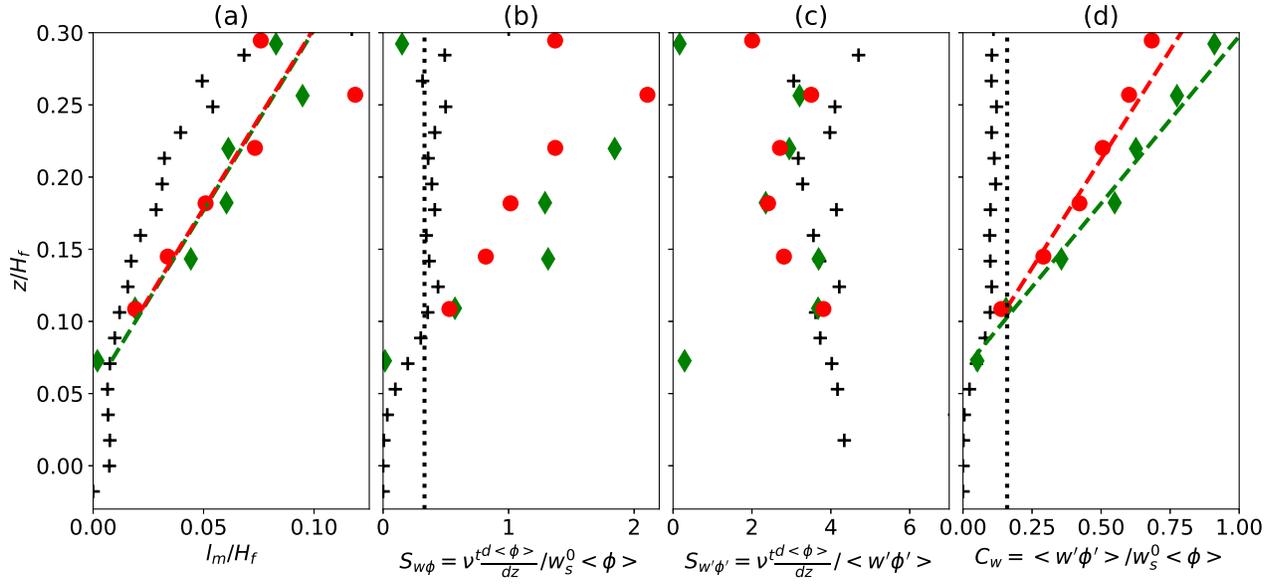}
}
\caption{Comparison of RB15 data (black +)  with $SL99\_1$ (green diamond $\diamond$) and $SL99\_2$ (red circle $\circ$)  data from \citet{shen1999}. Same panels as figure  \ref{fig2}. Dashed lines in panel (d) represents linear fit to the data.}
\label{fig4}
\end{figure*}

To further support our hypothesis, the measured data reported by \citet{shen1999} are further examined using the same methodology as described above. The hydrodynamic conditions of these two runs (red and green symbols in Figures \ref{fig3} and \ref{fig4}) are very similar to RB15 flow conditions  (see figure \ref{fig3} and table \ref{table_run_properties}), the major difference concerns the particles properties that are denser and smaller, leading to smaller inertia (see section II.A), with a suspended particle concentration two orders of magnitude smaller. 

Figure \ref{fig4}a shows the mixing length profiles with a von K\'arm\'an constant value very close to 0.41. This suggests that for lower suspended particle concentration, no reduction of the von K\'arm\'an constant is observed. In figure \ref{fig4}b, the turbulent Schmidt number $S_{w\phi}$ shows a similar trend as for RB15 in the near wall region where the value is lower than unity but the scatter in \citet{shen1999} increases away from the wall where $S_{w_\phi}$ exceeds unity. The vertical profiles of $S_{w'\phi'}$ are shown in figure \ref{fig4}c. Almost all data reveal values well above unity supporting lower turbulent particle concentration diffusivity compared to eddy-viscosity. Finally, the vertical profile of the dimensionless settling velocity ${\rm C_w}$ is shown in figure \ref{fig4}d. Both runs exhibit values smaller than unity confirming the hypothesis of settling retardation but with a weaker magnitude. Furthermore, the ${\rm C_w}$ profiles increase linearly with distance from the bed whereas the RB15 data exhibit a fairly constant value. 

\begin{figure*}[b!]
\centerline{
\includegraphics[width=\textwidth]{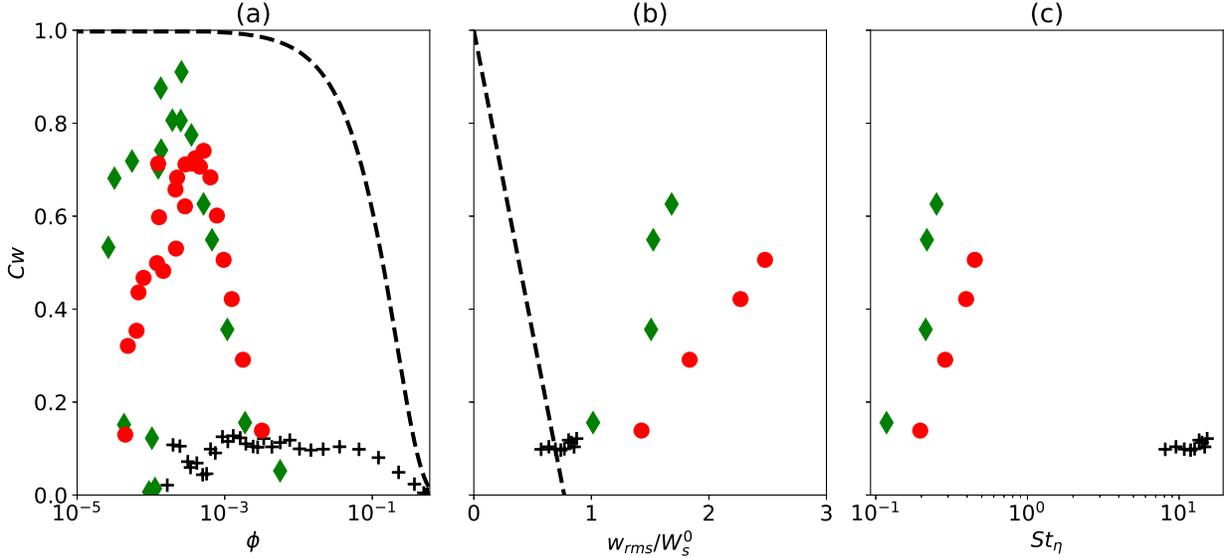}
}
\caption{Evaluation of different settling retardation mechanisms in RB15 data (black +) ,  SL99\_1 (red circle $\circ$) and SL99\_2 (green diamond $\diamond$) data. (a) Hindered settling function ${\rm C_w}=f(\phi)$, the black dashed line represents the \citet{richardson1954} function (${\rm C_w}=(1-\phi)^{4.65}$) ; (b) Scaling proposed by \citet{akutina2020} ${\rm C_w}=f(W_{rms}/W^0_s)$ and empirical law model proposed by \citet{akutina2020} (Eq. \ref{EffectiveSettling}) ; (c) Scaling with the Stokes number ${\rm C_w}=f({\rm St_\eta})$ where ${\rm St_\eta}=\tau_p/\tau_\eta$ with $\tau_p$  the particle response time and $\tau_\eta=(\nu^f/\epsilon)^{1/2}$ and $\epsilon = W_{rms}^3/d_p$.}
\label{fig5}
\end{figure*}

Both experimental datasets  \citet{revil-baudard2015,shen1999} and the two-fluid LES \cite{cheng2018} support the conclusion that the actual turbulent Schmidt number is higher than unity and that settling retardation plays an important role in sediment transport. In the following, we will use existing data to strengthen the analysis on settling retardation in turbulent flow. In figure \ref{fig5}a the dimensionless settling velocity ${\rm C_w}$ is plotted as a function of the local volumetric concentration in semi-log scale to test the hypothesis of hindrance effects. The dashed line in the figure represents the classical \citet{richardson1954} empirical formula ${\rm C_w}=(1-\phi)^{4.65}$. Clearly, the observed reduction of settling velocity can not be explained by hindrance effect. Indeed settling reduction occurs for volumetric concentration as low as $10^{-3}$ for \citet{shen1999} data and there is almost no variation with $\phi$ in RB15 data. In figure \ref{fig5}b the dimensionless settling velocity ${\rm C_w}$ is plotted as a function of $W_{rms}/W_s^0$ as suggested by \citet{akutina2020}. As already pointed out, the scaling proposed by  \citet{akutina2020} is compatible with RB15 data, however experiments by \citet{shen1999} corresponds to $W_{rms}/W_s^0>1$ which are out of  \citet{akutina2020} experimental range. It would be interesting to test this range of $W_{rms}/W_s^0$ in the settling column to conclude on the observed trends in \citet{shen1999} data but this is beyond the scope of the present contribution. In figure \ref{fig5}c, the dimensionless settling velocity ${\rm C_w}$ is plotted as a function of ${\rm St_\eta}$ where the Kolmogorov time-scale has been evaluated as $\tau_\eta = (\nu^f/\epsilon)^{1/2}$ with $\epsilon = W_{rms}^3/d_p$. The choice of $W_{rms}^3$ is justified by the anisotropy of the turbulence in the near-bed region and the focus on the wall-normal particle-flux that is driven by wall-normal velocity fluctuations. Using this local estimation, the Stokes number shows that it is lower than unity for \citet{shen1999} and higher than unity for \citet{revil-baudard2015} data (the data in this panel are only plotted in the region $0.3>z/H_f>0.1$). Therefore, the observed differences between the two datasets may be attributed to the differences in particles inertia but this would deserve further investigations based on experimental data over a wider range of flow conditions and particle properties.
 
\begin{figure*}[t]
\centerline{
\includegraphics[width=\textwidth]{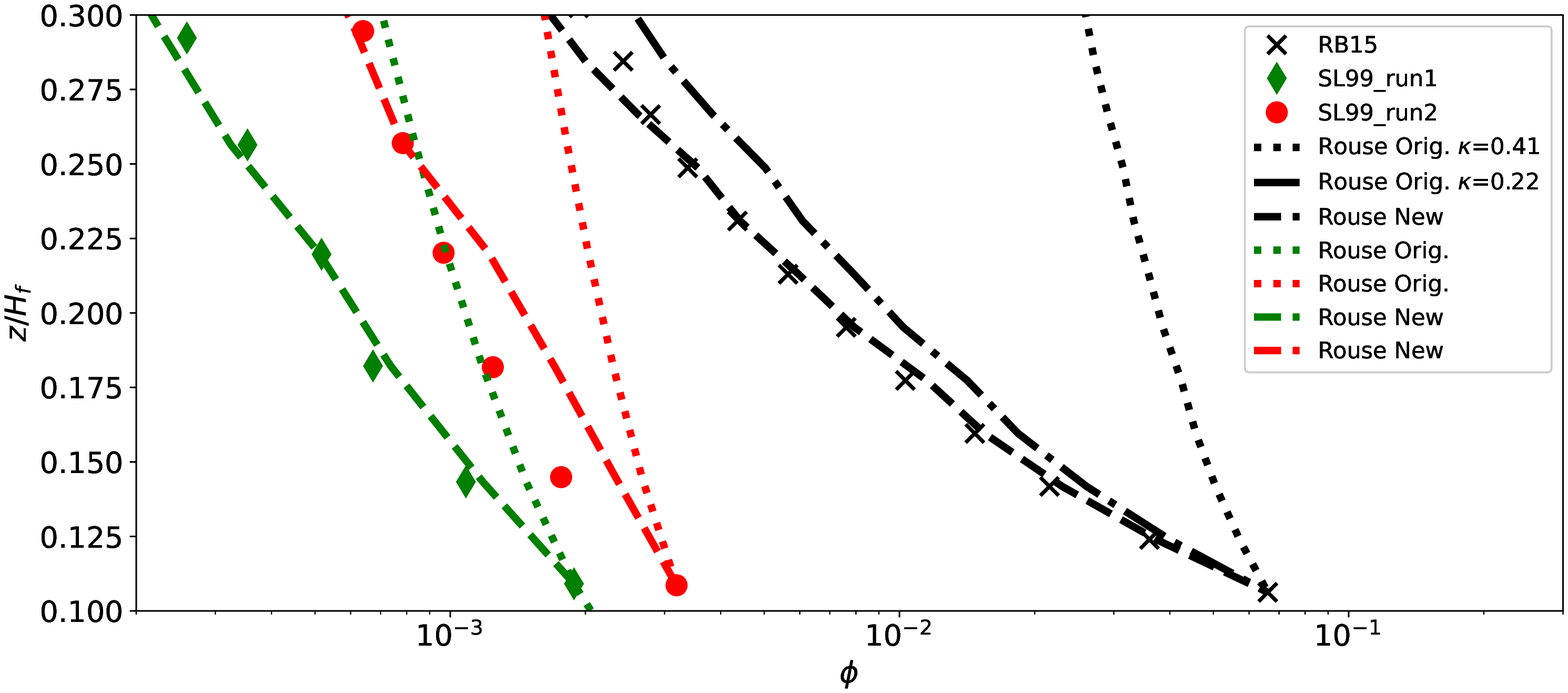}
}
\caption{Comparison of Rouse profile predictions with RB15 data (black +) ,  SL99\_1 (red circle $\circ$) and SL99\_2 (green diamond $\diamond$) data. Dotted lines represents the solution of Eq. (\ref{RouseProfile}) with $\kappa=0.4$ and the value of ${\rm S_c=S_{w\phi}}$ given in table \ref{table_Rouseparams} denoted as Rouse Orig. The dash-dotted line corresponds to the Rouse Orig. formula using $\kappa=0.22$ for RB15. Dashed lines represents  the solution of Eq. (\ref{RouseIntegrated}) with depth-averaged values given in table \ref{table_Rouseparams} and ${\rm S_c=S_{w^\prime \phi^\prime}}$. }
\label{fig6}
\end{figure*}

In order to evaluate the consequences of settling retardation and turbulent Schmidt number higher than unity on volumetric concentration profile prediction, Eq. (\ref{MassBalance})  is integrated numerically from $z_r/H_f=0.1$ to $z/H_f$:
\begin{equation}
\label{RouseIntegrated}
\displaystyle \langle \phi \rangle (z)= \int_{z_r}^z \dfrac{{\rm S_c} \ {\rm C_w} \ W^0_s}{\kappa^2 \ (z-z_d)^2\big\vert  {\rm d} \langle U\rangle /  {\rm d} z\big\vert} dz.
\end{equation}
Using ${\rm S_c}=S_{w\phi}$, ${\rm C_w}=1$ and $\kappa=0.4$ the result of Eq. (\ref{RouseIntegrated}) is denoted as Rouse Orig in figure \ref{fig6} while using  ${\rm S_c}=S_{w^\prime \phi^\prime}$ and measured ${\rm C_w}$ (using fit from figure \ref{fig4} for SL99\_1 and 2) the result of Eq. (\ref{RouseIntegrated}) is denoted as Rouse New in figure \ref{fig6}. For RB15, the dash dotted line is obtained by using $S_{w\phi}$ and ${\rm C_w}=1$ but with $\kappa=0.22$ as deduced from the fit in figure \ref{fig2}. The original Rouse formulation provides poor predictions of concentration profiles for all cases. The correction provided by $S_{w\phi}$ fails to predict the slope of the concentration profiles for SL99\_1 and SL99\_2. For RB15 data with high sediment concentration, the modification of the von K\'arm\'an constant has a significant impact on the prediction illustrating the importance of accounting for turbulence modifications, \textit{i.e.} modification of $\kappa$, in the prediction of suspended particle concentration. For all cases, using a turbulent Schmidt number $S_{w^\prime \phi^\prime}=3.75$ and a dimensionless settling velocity ${\rm C_w}$ provides the best prediction. Nevertheless, the rather crude resolution in SL99\_1 and 2 data and the limited range of parameters investigated herein is not sufficient to give definitive conclusion on the parametrization for the Rouse profile. 

\begin{table}[ht]
\caption{Summary of Rouse profile parameters. The turbulent Schmidt numbers and dimensionless settling velocity are averaged vertically over the range $0.25>z/H_f>0.1$ for all configurations except ${\rm C_w}$ for SL99\_1 and 2 for which a linear fit has been performed.}
\begin{center}
\begin{tabular}{cccccccccccccc}
\hline
\hline
Name            &           Reference                & $\kappa$ & $z_d/H_f$ & $S_{w\phi}$    & $S_{w^\prime \phi^\prime}$ & ${\rm C_w}$  \\
\hline
$RB15$     &\citet{revil-baudard2015}      & $0.22$     &    0.046      &  $0.38$            & $3.75$                                 & $0.1$       \\
$SL99\_1$ &\citet{shen1999}                   & $0.4$       &    0.052      & $1.06$             & $3.24$                                 & $4.3\times(z/H_f-0.066)$        \\
$SL99\_2$ &\citet{shen1999}                   & $0.4$       &    0.053     & $0.79$             & $3.01$                                 & $3.3\times(z/H_f-0.061)$        \\
 \hline 
 \hline
\end{tabular}
\end{center}
\label{table_Rouseparams}
\end{table}


\section{Conclusion}

In summary, the results presented herein reveals that the uncertainties on the value of the turbulent Schmidt number can be explained by a shortcoming in the methodology, and the real value of the turbulent Schmidt number estimated from directly measured Reynolds particle flux is larger than unity, ${\rm S_c} \in [3; 4]$ and that the settling velocity is significantly reduced ${\rm C_w} \in [0.1; 0.9]$ even at very low particle concentration. In the experiments with "large" plastic particles, the dimensionless settling velocity is almost constant in the near bed region, however, for fine sand particles, the settling velocity is linearly increasing with the distance from the bed. A comparison with existing models reveals that the observed settling reduction can not be explained by hindrance effects, while turbulence-particle interactions are most probably responsible for this behavior. Another important result  presented here concerns the modification of turbulence in the boundary layer and its consequences on suspended particle profile. In \citet{revil-baudard2015} measurements and two-fluid LES the von K\'arm\'an constant is reduced by a factor two. The consequences of this turbulence modification on particle concentration profile are huge and definitely needs to be accounted for in the Rouse profile parametrization.

In conclusion, innovative highly-resolved measurement technique and numerical simulations allow us to obtain new insight that changes the paradigm of suspended particle transport in a boundary layer flow. We have demonstrated that (i) turbulent dispersion of inertial particles is significantly smaller than that of fluid parcels and does not depend much on particle size and density and (ii) settling velocity of inertial particles is reduced and this reduction depends on particle size and density. This suggests that the particle properties are sensitive parameters in this problem but the details of the underlying physical mechanisms are still to be elucidated. Beyond the importance for sediment transport predictions, these findings are relevant to a wide range of two-phase flows such as avalanches, turbidity currents, pneumatic transport and particle deposition in material processing.

\section*{Data availability }

The data and post-processing script used to write the manuscript are available on zenodo at the following address \url{https://zenodo.org/record/5765565#.Ya_L7VPjJHt} (see reference \cite{chauchatPRFData}).

%
%

\begin{acknowledgments}

J. Chauchat is financially supported by the french ANR project SHEET-FLOW (ANR-18-CE01-0003). D. Hurther is supported by the french DGA-funded ANR Astrid Maturation project MESURE (ANR-16-ASMA-0005). Z. Cheng and T.-J. Hsu are supported by U.S. Strategic Environmental Research and Development Program (SERDP, MR20-1478) and National Science Foundation (OCE-1635151). The computations presented in this paper were performed using the GENCI infrastructure under Allocations A0060107567 and A0080107567 and the GRICAD infrastructure.

The authors would like to thank N. Mordant and P. Frey for their constructive comments on the manuscript.

\end{acknowledgments}




%
%
\bibliography{bibliographie_Bibdesk}

\end{document}